\documentclass[twocolumn,apj]{aastex63}
\usepackage{amsmath,amssymb,gensymb,color}
\usepackage{natbib,hyperref}
\usepackage[utf8]{inputenc}
\usepackage[T1]{fontenc}

\defcitealias{p9-messenger}{S23}

\begin{document}

\title{Meteor CNEOS 2014-01-08 has nothing to do with Planet 9}
\author[0000-0002-4478-7111]{Sigurd~Naess}
\affil{Institute for theoretical astrophysics, University of Oslo, Norway}

\begin{abstract}
It has been suggested that a gravitational slingshot from the hypothetical Planet 9 (P9)
could explain the unusually large velocity of meteor CNEOS 2014-01-08. I show that
this explanation does not work because P9 can at most provide an insignificant
0.25 km/s of the object's 42 km/s asymptotic heliocentric velocity and at most
a 7.6 degree deflection due to P9's low orbital speed and non-zero radius.
Furthermore, the hypothesis requires an encounter with two planets that is
trillions of times more unlikely than CNEOS 2014-01-08 simply being
fast from the beginning.
\end{abstract}

\section{The Messenger Hypothesis}
In a search of the CNEOS meteor catalog, \citet{cneos-interstellar}
identified meteor CNEOS 2014-01-08 (C14 hereafter) as likely interstellar object, with
an impact speed of 44.8 km/s, a heliocentric speed at infinity of $v_{\textrm{C14},\infty}=42.1\pm5.5$ km/s,
and a speed of $v_\textrm{C14,LSR}=58\pm6$ km/s relative to the local standard of rest (LSR) in the Sun's
neighborhood in the galaxy.\footnote{It's unclear where the uncertainties
in \citet{cneos-interstellar} come from, since the CNEOS catalog does not
contain uncertainties.} \citet{p9-messenger} (S23 hereafter) argues that C14's
trajectory is suspicious for two main reasons:

\begin{enumerate}
\item Its high $v_\textrm{C14,LSR}$ is a $1.5\sigma$ ``outlier'' given the
38 km/s velocity dispersion of stars in the solar neighborhood. This is
mildly unlikely -- 6.3\% of stars should move this fast or faster assuming
a Gaussian distribution. The actual distribution is non-Gaussian though,
with different star populations having different distributions, and
objects ejected from these stellar systems would be expected to be somewhat
faster than these.
\item Its extrapolated trajectory passes through the parts of the sky
where the hypothetical Planet 9 (P9 hereafter) would be most likely to be, if it
exists \citep{p9-orbit} -- a 1-9\% coincidence.
\end{enumerate}

I do not find these probabilities suspiciously small myself, but
\citetalias{p9-messenger} suggest that C14's high speed and
direction of origin could be explained if it had been deflected and
accelerated towards Earth by a gravitational slingshot with P9.
They call this the ``messenger'' hypothesis because C14 would tell us
where to look for P9, and it recently motivated an optical search
for P9 in the direction C14 came from \citep{p9-messenger-search}.
It's an interesting hypothesis, but as I will show it does not work in practice.

\section{A Planet 9 gravitational slingshot}
Let us consider how a P9 gravitational slingshot would affect C14.
Without loss of generality, we can work in two dimensions and place
the Sun (with the nearby Earth) at the origin, and P9 at
coordinates $[0,r_9]$. P9's distance from the Sun $r_9$ is
unknown, but likely in the range $250-700$ AU. P9's orbit
is only expected to be moderately eccentric, which we will
approximate as circular here, giving it a velocity of
$\vec v_9 = [v_9,0]$, with
\begin{align}
	v_9 &= 1.9 \textrm{km}/\textrm{s} / \sqrt{r/{250\textrm{AU}}}.
\end{align}
So the closest, and therefore fastest, version of P9 would be
expected to move at only around 2 km/s.

In a gravitational slingshot, C14 has an incoming velocity
$\vec v_1 = [v_{1x},v_{1y}]$ and outgoing velocity $\vec v_2 = [v_{2x},v_{2y}]$.
It is much less massive than P9, so we can take P9's orbit to be
unaffected. Then, by conservation of momentum, C14's speed in
P9's reference frame must be unaffected by the deflection:
\begin{align}
	|\vec v_1-\vec v_9| &= |\vec v_2-\vec v_9| \Leftrightarrow \notag \\
	v_1^2 - 2v_{1x}v_9 &= v_2^2 - 2v_{2x}v_9
\end{align}
C14 must be deflected towards the inner solar system, so we must have
$v_{2x}=0$. Introducing the deflection angle $\alpha$, we get
\begin{align}
	v_2^2 &= v_1(v_1+2\sin\alpha v_9) \notag \\
	v_1 &= \sqrt{v_2^2 + v_9^2\sin\alpha^2} - v_9\sin\alpha \approx v_2 - v_9\sin\alpha
\end{align}
where the approximation is good to 0.1\% error for $v_2 = v_{\textrm{C14},\infty} = 42$ km/s like
for C14. So the speed gain is at most simply P9's orbital speed, so up to 1.9 km/s.

This may seem too small to be relevant, and indeed in terms of heliocentric speed
a P9 slingshot doesn't do much to C14. What is useful is not the speed change itself, but
the redirection of C14's trajectory, which can change it from moving opposite
to the local standard of rest to moving along it. According to \citet{cneos-interstellar},
C14's asymptotic heliocentric velocity is $\vec v_{\textrm{C14},\infty} = [32.7,-4.5,26.1]$ km/s
in galactic coordinates, compared to the local standard of rest's
$\vec v_\textrm{LSR} = [-11.1,-12.2,-7.3]$ km/s. If C14's velocity could
be redirected along the LSR, its LSR-relative speed would fall from
65 km/s to 24 km/s, even without any change in heliocentric speed. This
would accomplish the messenger hypothesis' goal of making C14 more normal
compared to the LSR.

\section{Problem 1: Planet 9 is too big}
\label{sec:big}
A strong gravitational field is needed to deflect a fast-moving object.
C14's closest distance to P9 is
\begin{align}
	r_\textrm{min} &= -a(e-1).
\end{align}
Here $a = -GM_9/v_\textrm{rel}^2$ is the hyperbolic equivalent of the semimajor axis,
and $e = -1/\cos(\beta/2) = 1/\sin(\alpha/2)$ is the eccentricity. $\beta$ is the external angle between the two
asymptotes of C14's path in P9's reference frame, so $\beta = \pi+\alpha$. $v_\textrm{rel}$
is C14's asymptotic speed relative to P9, and is simply given by
\begin{align}
	v_\textrm{rel} &= \sqrt{v_2^2 + v_9^2} \approx v_2 = 42 \textrm{km/s}
\end{align}
Of course, C14 can't get arbitrarily close to P9 without hitting the planet itself.\footnote{
There is also the Roche limit, which would be even more limiting, but which may not apply
to small objects like C14 that can have significant structural strength.} This means that
the larger P9's radius, the less suitable it is for gravitational slingshots, all other things equal.
P9 is expected to have a mass of roughly $M_9 \approx 6 M_\Earth$ and a radius of
roughly $R_9 \approx 3 R_\Earth$ \citep{p9-orbit,p9-atmosphere-2016}. We can use this
to infer the allowed range of deflection angles.
\begin{align}
	e-1 > 14 \cdot \Big(\frac{R_9}{3R_\earth}\Big) \Big(\frac{M_9}{6 M_\earth}\Big)^{-1}, \label{eq:e}
\end{align}
which translates to a maximum deflection angle of
\begin{align}
\alpha &= 2\sin^{-1}(1/e) < 7.6\degree
\end{align}
for the P9 mass and radius chosen above. Hence, P9 could not meaningfully deflect
C14 even under optimal circumstances. This not only prevents P9 from aligning it
with the local standard of rest, it also further limits the speed gain to a
paltry
\begin{align}
\Delta v \approx v_9\sin\alpha < 0.25 \textrm{km/s}
\end{align}
This limit could be avoided if P9 is much more compact than expected
(e.g. a primordial black hole has been suggested \citep{p9-black-hole}),
though in my opinion this is stacking hypotheticals much too deeply
given the flimsy evidence for both P9 and primordial black holes.

\section{Problem 2: Double encounters are very unlikely}
Both the messenger hypothesis and the null hypothesis require C14 to have
a close encounter with a planet (P9 and Earth respectively). But the messenger
hypothesis additionally requires C14 to go on to be deflected onto an intersection
course with a \emph{second} planet (Earth). We can estimate the probability of
this happening as the fraction of the sky covered by Earth as seen from P9, so
\begin{align}
P_{9\rightarrow\earth} &\approx \frac{\pi R_\earth^2}{4\pi r_9^2}
	= 7\cdot 10^{-15} \Big(\frac{r_9}{250 \textrm{AU}}\Big)^{-2}.
\end{align}
Put another way, for every C14 that's deflected by P9 to an Earth impact,
another $10^{14}$ such objects would have had a close encounter
with P9 but deflected in the wrong direction.

If we define P9's ``deflection radius''
$\rho$ as the radius at which it can deflect C14 by at least 90\degree
(so $e = \sqrt{2}$), then by equation \ref{eq:e} we get
\begin{align}
	\rho &= 0.089 R_\earth \cdot \frac{M_9}{6M_\earth}
\end{align}
This is much smaller than P9 is expected to be, but for the sake of
the argument, let's assume it's compact enough that we don't have to
worry about its surface. Then we find that to be strongly deflected
by P9, a C14-like object must hit a target with a cross section around
100 times smaller than Earth.

Assuming the density of interstellar objects is not dramatically
different in the inner vs. outer solar system, this implies that for
every C14 that's deflected by P9 to hit Earth, we expect around
$10^{16}$ such objects should hit Earth directly! Given the $\sim$
10-year coverage of CNEOS, this works out to around 1 quadrillion
C14-like meteors per year, which is clearly absurd. C14 had a mass
of around 500 kg, so this would add around 100 $M_\earth$ to Earth
per billion years and deposit a power of $\sim 10^{19}$ W, rising
its temperature to around 1000 K. I think it's safe to say this
isn't happening.

\section{Conclusion}
\citet{p9-messenger} argues that a gravitational slingshot with Planet 9
avoids the problem of a surprisingly high population of high-velocity
interstellar minor objects. I have shown that not only is Planet 9 too slow
to change their speed much and too big to change their direction meaningfully,
but even if it could, it would require such an unlikely deflection that it
the implied population of high-velocity interstellar objects would be
trillions of times higher than under the null hypothesis. The messenger
hypothesis was an interesting idea, but it falls apart when one actually
looks at the numbers.

\bibliographystyle{act_titles}
\bibliography{refs}

\end{document}